# Mobile phone enabled Supply chain management in the RMG sector: A conceptual framework

## Dr. Md. Taimur Ahad

Assistant Professor
Department of Computer Science
Faculty of Engineering and Technology
Eastern University, Bangladesh

*Ph.D. in Computer Science (Macquarie University)*
*M. Sc. in Computing Science (University of Technology Sydney)*
*Masters in IT (University of Western Sydney)*

Email: taimur.cse@eastermuni.edu.bd

## Abstract

Relatively little is known about mobile phone use in a Supply Chain Management (SCM) context, especially in the Bangladeshi Ready-Made Garment (RMG) industry. RMG is a very important industry for the Bangladeshi economy, but is criticized for long product supply times due to poor SCM. RMG requires obtaining real-time information and enhanced dynamic control, through utilizing information sharing and connecting stakeholders in garment manufacture. However, a lack of IT support in the Bangladeshi RMG sector, the high price of computers and the low level of adoption of the computer based internet are obstacles to providing sophisticated computer aided SCM. Alternatively, explosive adoption of mobile phones and continuous improvement of this technology is an opportunity to provide mobile based SCM for the RMG sector. This research presents a mobile phone based SCM framework for the Bangladeshi RMG sector. The proposed framework shows that mobile phone based SCM can positively impact communication, information exchange, information retrieval and flow, coordination and management, which represent the main processes of effective SCM. However, to capitalize on these benefits, it is also important to discover the critical success factors and barriers to mobile SCM systems.

## Keywords

Mobile phone based supply chain, innovative SCM, Supply chain in manufacturing, SC and technology.

## Introduction

In the information age, successful Supply Chain Management (SCM) requires information exchange and integration within the SCM network. Information availability and flow integrates all key business activities through improved relationships at all levels of the SCM (internal operations, upstream supplier networks and downstream distribution channels (Al-Mashari&Zairi 2000). Mobile phone technologies, by virtue of being carriers and conduits of information, play a role in providing large-scale information transfer within SC (Abraham 2006; Doolin et al. 2008; Gonzálvez-Gallego et al. 2015; Kim et al. 2015; Kumar et al. 2014; Liang 2015; Qrunfleh&Tarafdar 2014; Singh &Garg 2015; Simatupang&Sridharan 2005; Tserng et al. 2005; Wang et al. 2007; Youn et al. 2014). The literature extols the potential benefits of Mobile phone integration within the SC and the crucial role of integrated IT to deliver those benefits and how information errors affect SC performance (Kwak&Gavirneni 2014). However, Kumar et al. (2015) advocates that the new IT technologies could pose new challenges in developing countries due to a lack of resources and directions available.

Recently, the term SCM has risen to prominence in mobile phone based research. Studies suggest implementing mobile phones may overcome information availability challenges, especially in resource scarce environments (Abraham 2006; Qrunfleh&Tarafdar 2014; Tserng et al. 2005; Wang et al. 2007). Isaksson et al. (2010) suggests that process innovation using mobile phones for SCM improves communication effectiveness for the SC. Furthermore Abraham (2006), Tserng et al. (2005) and Wang et al. (2007) conclude that communication and information availability are the main benefits of mobile-based SC. However, as environmental factors and social structures also play an important role in SC, Isaksson et al. (2010) suggests it is essential to study the critical success factors of mobile based SC.

Another motivation for undertaking this study is that very little research has been undertaken on the Bangladeshi RMG sector with regard to capitalizing on the benefits of ICT for the effectiveness of SC. This lack of consideration partly explains why the Bangladeshi RMG sector finds it difficult to address the issue of long product lead-times. Given the aim of investigating SC performance on Bangladeshi RMGs, this study presents the literature as well as a conceptual mobile phone based SCM framework for the RMG sector in Bangladesh. The paper concludes with questions for future research.

## Literature review

*1.1 Mobile technology usage in SCM*

With the characteristics of a mobile phone and its enabling technologies such as network technologies, service technologies, mobile middleware, mobile commerce terminals, mobile location technologies, mobile personalization technologies, and content delivery and format - has the potential to make information flows more efficient and coordinate the operations within the extended enterprise (Siau&Shen 2002). Mobile technology applications such as mobile email and internet for corporate users, mobile customer care and mobile enterprise implementations, represent some enabling technologies that can be deployed for SCM (Siau&Shen 2002).

The main potential of mobile phones is in improving communication. A mobile phone can support internal, external and inter-organizational communication (Litan&Rivlin 2001). Mobile phone communication provides improved communication between firms and their suppliers, mobile phones can enable firms to manage their SCs more effectively, streamline their production processes and engage in new activities (Hardy, 1980; Roller and Waverman, 2001).

Doolin& Ali (2008) focused mobile technology in the SC and reported the most important factors influencing mobile phone adoption were technological innovation and information intensity (transfer and use) of the company. Other factors that appeared to influence mobile adoption included the compatibility of the technology with the company's business approach, the presence of top management support, and the degree of organizational readiness. Environmental factors such as competition within the industry or business partner influence seemed less influential for these pioneers of mobile technology use in supply-side activities (Doolin& Ali 2008).

Wang et al. (2007) implemented a mobile RFID-based SC control and found that information sharing can reduce project conflicts and project delay. Improved communication, convenient information sharing, ease of data/information acquisition, reliability, accuracy and comprehensive functionality represent some of the reported benefits of implementing mobile based SC. Other studies by Cagliano et al. (2015), Nair et al. (2015) and Tserng et al. (2005) also suggested mobile phone can be extremely useful in improving the effectiveness and convenience of information flow in SCM. Abraham (2006) reported evidence from the fishing industry in India that markets became more efficient with the introduction of mobile phones with subsequent freer flow of information. Key advantages for the fishing industry here were price dispersion and fluctuation, less wastage of time and resources and reduced risk and uncertainty (Abraham 2006).

*1.2 Critical success factor (CSF) for effective SCM*

Previous SCM studies identified several CSFs that ensures SC to flourish and goals to be attained. The identified CSFs are listed in Table 1.

The SC is a network of businesses, it is therefore important to understand and examine how the SC network structure is configured (Lambert & Cooper 2000). SCM therefore increasingly is recognized as the management of key business processes across the network of organizations that comprise the SC (Choon et al. 2002; Cooper et al. 1997; Lambert & Cooper 2000).

Communication has been identified as the one of the main CSFs in establishing effective SCM (Chen et al., 2004; Denolf et al. 2015; Prahinksi and Benton 2004). Appropriate communication strategies enhance supplier firm's operational performance (Chen et al., 2004; Denolf et al. 2015; Prahinksi and Benton 2004). Effective and efficient communication between SC partners reduces product cost and performance-related errors, thereby enhancing quality, time, and customer responsiveness (Carr and Pearson 1999; Chen and Paulraj 2004; Dyer 1996 ;Paulraj et al. 2008;Turnbull et al., 1992).

Table 1: Main critical success factors (CFS) and barriers identified in previous SCM studies.

| Success factors/Barriers | Study |
| --- | --- |
| Communication | Chan et al. (2004), Denolf et al. (2015) |
| Management of business process | Lambert & Cooper (2000), Wang et al. (2007) |
| Integration of chains | Choon et al. (2002), Mentzer et al (2001), Prajogo&Olhager (2012) Simatupang&Sridharan (2005) |
| Collaborative performance | Lee et al (2014) |
| Competitive advantage | Li et al.(2006), Doolin& Ali (2008), Kumar et al. (2015) |
| Better SC project management | Akintoye et a. (2000), Abraham (2006), Chan et al. (2004), Denolf et al. (2015), Lee et al (2014) |
| Top management support | Akintoye et a. (2000), Kumar et al. (2015), Doolin& Ali (2008) Lee et al (2014) |
| Decision synchronization | Simatupang&Sridharan (2005) |
| Information flow and sharing | Choon et al. (2002), Mentzer et al (2001), Simatupang&Sridharan (2005), Tserng et al. (2005), Lee et al (2014), Prajogo&Olhager (2012) |
| IT Alignment and strategy | Tserng et al. (2005), Harland et al. 2007, Youn et al. (2014), Lee et al (2014), Denolf et al. (2015), Doolin& Ali (2008), Prajogo&Olhager (2012), Qrunfleh&Tarafdar (2014), Wittstruck&Teuteberg (2012) |
| Customer service management | Choon et al. (2002) |
| Higher levels of SCM practice | Li et al. (2006) |

Simatupang&Sridharan (2005) emphasized collaborative performance of the SC partners and the interaction phenomena among different features of the SC. Their study suggests that collaborative systems, information sharing, decision synchronization, incentive alignment and integrated SC processes are necessary to improving SC. Recently Özdemir et al. (2015) also suggested supply chain integration as one of the CSFs in Turkish small business SCM. Li et al.'s (2006) study reported higher levels of SCM practice could lead to enhanced competitive advantage and improved organizational performance of the SC. Prajogo&Olhager (2012) also examined the role of integration and the study reported logistics integration had a significant effect on operations performance.

Prajogo&Olhager (2012) examined the role of the long-term supplier have both significant direct and indirect effects on SCM performance. Chan et al. (2004) study indicated the establishment and communication of a conflict resolution strategy, a willingness to share resources among project participants, a clear definition of responsibilities, a commitment to a win-win attitude, and regular monitoring of partnering process - were the significant underlying factors for SC collaboration success. Wittstruck&Teuteberg (2012) takes a sustainability viewpoint on SC and found that signalling, information provision and the adoption of standards are crucial preconditions for strategy commitment, mutual learning hence the overall success of SCM. Improved production planning and purchasing (Akintoye et al. 2000), customer service management (Choon et al. 2002) were also identified as CSFs in previous SCM research studies.

*1.3    Barriers in Effective SC*

Serdarasan (2013) identified three types of interrelated SC complexity- Static complexity describes the structure of the SC, the variety of its components and strengths of interactions; dynamic complexity represents the operational uncertainty in the SC and involves aspects of time and randomness; and decision-making complexity involves the volume and nature of the information considered when making a SC related decision. Harland et al. (2007) suggested that not only the lack of information, but also the lack of SC alignment to IS strategy is another barrier to SC information integration. Qrunfleh&Tarafdar (2014) also emphasized that SC information system strategy largely impacts on SC performance and firm performance. Among other barriers discussed in previous literature, include the lack of strategic alignment of information strategies in the SC, firm size of some SC actors, lack of awareness of the potential benefits of ICT, lack of motivation, and being in a less developed industry or regional context (Harland et al. 2007). Ability to adopt IT, the high cost of ICT products, specific organizational needs, inappropriate business size for ICT, are also some of the issues relating to barriers in the uptake of IT in SCM (Kumar et al. 2015). Akintoye et al.'s (2000) study suggest barriers to SC success include: workplace culture, lack of senior management commitment, inappropriate support structures and a lack of knowledge of SCM philosophy. Training and education at all levels in the industry are necessary to overcome these barriers.

## 2 RESEARCH GAP AND QUESTIONS

Potentially mobile phone and mobile technology in general could improve SCM in Bangladeshi RMG. Earlier studies suggest the use of ICT and the development of IT based SC systems has been slower than expected, particularly in SMEs in developing countries (Harland et al. 2007; Kumar et al. 2015; Lee et al. 2014). A recent study by Denolf et al. (2015) argues analysis showed that some CSFs have been ignored and important supply chain characteristics have been overlooked in contemporary SCM studies. SCM models should be based on actual industry practices and SCM models should address how SC can assist the business (Choon et al. 2002). Lastly, in this regard Isaksson et al. (2010) makes another justification for this research, to consider the environmental factors and social factors as contextual and the social structures that play an important role in SC. In Bangladesh, RMG consists of SMEs, it is therefore suggested to study the dynamics of these small firms (Barrett &Rainnie 2002) and to examine the different variables (Edwards & Ram 2006). However, to understand the underpinning concept of this study, it is important to take an 'integrated' perspective of SCM in Bangladeshi RMGs. The research questions and our general conceptual framework (figure 1) are as follows:

1. How can mobile phone based SCM support information sharing in the Bangladeshi RMG?

2. Which CSFs explain the implementation of mobile based SCM within RMG in Bangladesh?

3. Which inhibitors limit mobile based SCM within RMG in Bangladesh?

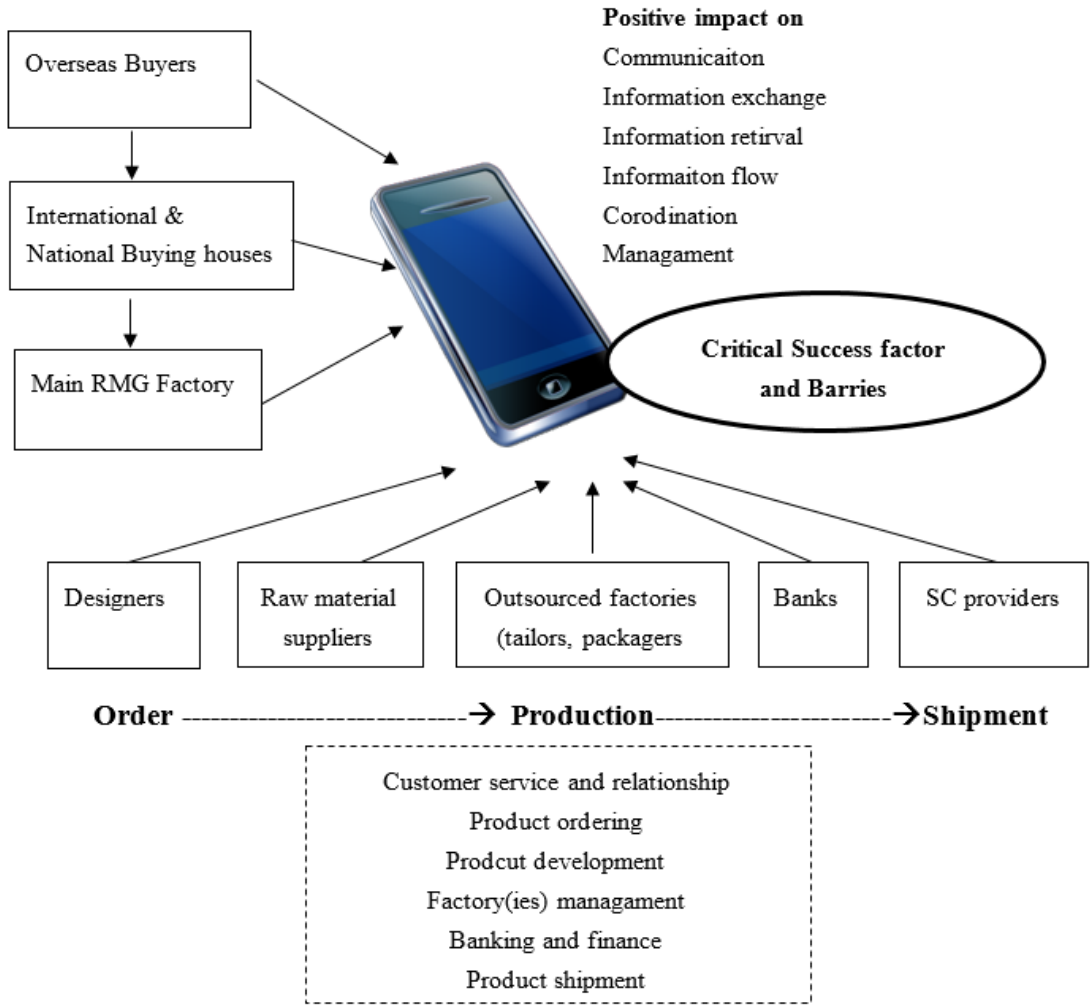

Figure 1. A conceptual framework on Mobile phone based SCM in Bangladesh RMG

## Future research direction and Conclusion

Future research should test the framework and extend the findings. More work is needed to explore the actual business practices in Bangladeshi RMG. Data should be collected from other parts of the industry to provide a greater perspective of SCM for Bangladeshi RMG. Future research should address SCM practices and evaluate the existing system. This research does not claim that all factors are identified, SC itself a complex process, so more exploratory investigation is required. However, the conceptual mobile based SC is an effort to navigate through the complex landscape of SC. The merit of the research lies in presenting for the first time a mobile phone based SC framework for RMG in Bangladesh, that considers customers and technological as well as business perspectives. Moreover, for the first time this research takes into considerationmobile phones, which are highly adopted, especially in developing counties. RMG is one of the main sources of income in the country. The Bangladeshi RMG is also important to rest of the world for producing quality affordable garments. Compared to other SC research, this research aims to understand the business practice of RMG. The contribution is significant, as it represents a step forward in mobile-based research.

*2.1.1*